\def\b{\bibitem}
\documentstyle[editedvolume,numreferences,psfig]{crckapb}

\begin{opening}
\title{Quantum Kinetic Theory: The Disordered Electron Problem}
\author{T.R.Kirkpatrick}
\institute{Institute for Physical Science and Technology, and Department of
         Physics\\
         University of Maryland, College Park, MD 20742}
\author{D.Belitz}
\institute{Department of Physics and Materials Science Institute,%\\
         University of Oregon,%\\
         Eugene, OR 97403}
\end{opening}
\begin{document}
\begin{abstract}
These are notes for lectures delivered at the NATO ASI on Dynamics in
Leiden, The Netherlands, in July 1998. The quantum kinetic theory for
noninteracting electrons in a disordered solid is introduced and discussed.
We first use many-body theory to derive the quantum Boltzmann
equation that describes transport and time correlation functions in this
system. Particular attention is paid to the calculation of the electrical
conductivity $\sigma $, and the density response function $\chi_{nn}$.
We then consider corrections to the Boltzmann equation due to wave
interference effects.
The disorder expansion of the conductivity is addressed, and the so-called
weak localization or long-time tail contribution to $\sigma $ is discussed.
We conclude with a brief discussion of the influence of
electron-electron interactions on the properties of
disordered electronic systems.

\vskip 25mm
 
\noindent
Contribution to:\quad  {\it Dynamics: Models and Kinetic Methods for Non-}
\par\noindent
\hskip 51mm {\it equilibrium Many-Body Systems}
\par\noindent
\hskip 32mm John Karkheck (editor)
\par\noindent
\hskip 32mm To be published by Kluwer academic publishers b.v.
\bigskip
\par\noindent
\hskip 32mm Lectures delivered by T.R. Kirkpatrick
\end{abstract}
\vfill\eject

\section{Introduction}
\label{sec:I}

Quantum kinetic theory has a long and interesting history. Shortly after the
discovery of the Pauli exclusion principle, Sommerfeld applied it
to electrons in metals and thereby resolved the most flagrant discrepancies
between the observed thermal behavior of solids and the predictions
of the classical Drude model of transport \cite{AshcroftMermin}. After the
introduction of quantum mechanics, the new theory was incorporated
into the standard Boltzmann description of transport. 
Like the classical Boltzmann equation, this quantum kinetic theory, also
known as the Uhlenbeck-Uehling equation \cite{UU},
is valid only for systems that 
are dilute in a sense to be explained below. Unlike the classical
Boltzmann equation, it properly takes into
account the statistics of the particles (fermionic or bosonic) as well as 
the quantum mechanical nature of the scattering process. However, it
misses more subtle quantum mechanical effects. To get a feeling for
what is missing, let us consider some length scales in the problem.

In a classical gas there are two length scales that occur in
transport problems: The mean freee path $\ell$, and 
the linear size of the particles or the impurities, which we denote by $a$. 
The Boltzmann transport theory is valid if
the system is dilute in the sense that $a/\ell\ll 1$.
In the quantum description of transport there is an additional length scale,
namely the de Broglie wavelength 
$\lambda$ of the scattered particle. Here we are interested in electron 
transport phenomena at temperatures that are low compared to the Fermi
temperature, so the relevant length is $\lambda_{\rm F}$, the de Broglie 
wavelength at the Fermi surface. Due to approximations made in its derivation,
the quantum Boltzmann equation or Uhlenbeck-Uehling equation is only valid if 
both $a/{\ell }\ll 1$ and ${\lambda_{\rm F}}/{\ell}\ll 1$. The latter
expansion parameter indicates the relative importance of quantum
mechanical wave interference effects.

In the first part of these lectures we introduce the standard Edwards
model \cite{Edwards,AGD} for noninteracting electrons in a disordered 
solid, and the
two response functions that are most important for our purposes, viz.
the electrical conductivity $\sigma$, and the density susceptibility
$\chi_{nn}({\bf k},\omega)$. We then use simple kinetic theory concepts
in conjunction with the particle number conservation law
to show that the behavior of the latter at small frequencies $\omega$ and
small wavevectors ${\bf k}$ is dominated by the hydrodynamic density
diffusion mode. Simple arguments lead to an estimate for the value
of the diffusion coefficient $D$ in terms of the collision mean-free time
$\tau$. $\sigma$ and $D$ are related by an Einstein relation,
\begin{equation}
\sigma =e^{2}\,\frac{\partial n}{\partial\mu}\,D\quad , 
\label{eq:1.1}
\end{equation}
with $e$ the electron change, $n$ the electron number density,
and $\mu $ the chemical potential. For noninteracting electrons at zero
temperature, $\partial n/\partial \mu=N_{\rm F}$, with $N_{\rm F}$ 
the single particle
density of states at the Fermi surface. Following this, we discuss how to
derive the quantum Boltzmann equation and how to compute $\sigma$ from it.

As mentioned above, the
quantum Boltzmann equation is valid only in the `semi-classical' limit
where $\lambda_{\rm F}/\ell \rightarrow 0$. In the remaining parts of these
lectures we discuss some of the interesting phenomena associated with wave
interference that appear at higher order in an expansion in powers of 
$\sim\lambda_{\rm F}/\ell$. In particular, the
nonanalytic nature of the disorder expansion of $\sigma $ is discussed as
well as weak-localization, or long-time tail, contributions to $\sigma$. We
conclude with a few remarks on the combined effects of disorder and
electron-electron interactions.

\section{The model, and basic concepts}
\label{sec:II}

\subsection{The model}
\label{subsec:II.A}

We will be concerned with the electronic transport
properties of disordered solids. In order to keep things simple,
we will mainly consider a model that ignores the electron-electron
interaction. While this approximation has yielded many important insights,
one should keep in mind that in general it is not justified. We will come
back to this point at the end of the lectures. For now we consider a
noninteracting electron gas model consisting of spin $1/2$ fermions moving
in a static random potential $u({\bf x})$. Physically, $u({\bf x})$ is produced
by the impurities in the solid, and is therefore
fixed for a given system, but we will assume that it is meaningful to consider
an ensemble of systems and to compute averages of obervables over the 
random potential.\footnote[1]{To what extent this assumption is justified is 
 a very
 tricky question that we cannot get into. In principle one should calculate
 a whole distribution for every observable. Calculating the average is
 sufficient if this distribution is sharply peaked about its average.}
In the standard Edwards model \cite{Edwards,AGD}, $u({\bf x})$ is taken to have
zero mean, and to be Gaussian distributed with the variance given by,
\begin{equation}
\left\{u({\bf x})u({\bf y})\right\}_{\rm dis} = \frac{\delta ({\bf x}-{\bf y})}
     {2\pi N_{\rm F}\tau}\quad .  
\label{eq:2.1}
\end{equation}
Here $N_{\rm F}=\left( m/2\pi \right) (k_{\rm F}/\pi )^{d-2}$ is the free
electron density of states per spin at the Fermi surface 
(in $d=2,3$ dimensions), $\tau$ is the elastic mean-free time between 
collisions, $\{\ldots\}_{\rm dis}$ denotes the disorder average, $k_{\rm F}$ 
is the Fermi wavenumber, and $m$ is the electron mass. Throughout the 
lectures we use units so that $\hbar =k_{\rm B}=e=1$, with $\hbar$ Planck's
constant, $k_{\rm B}$ Boltzmann's constant, and $e$ minus the electron
charge.

To motivate this model we consider a quantum Lorentz gas, i.e. a system
of moving electrons that interact with $N$ stationary hard 
scatterers of radius $a$,
but not with one another. The disorder average consists of averaging over the 
positions of the $N$ scatterers in the system volume $V$, 
\footnote[2]{Here we neglect correlations between the scatterers, which
are not important for our purposes.}
%
%\begin{mathletters}
\label{eqs:2.2}
\begin{equation}
\left\{\left(\ldots\right)\right\}_{\rm dis}  = 
    \frac{1}{V^{N}}\int d{\bf R}_{1}\ldots d{\bf R}_{N}\ (\ldots)\quad,
\label{eq:2.2a}
\end{equation}
and the electron-impurity potential can be written,
\begin{equation}
v({\bf x})=\sum\limits_{i=1}^{N}\ v({\bf x}-{\bf R}_{i})\quad, 
\label{eq:2.2b}
\end{equation}
with $\left\{ {\bf R}_{i}\right\} $ the positions of the $N$ scatterers or
impurities. Assuming that $\lambda (\approx \lambda_{\rm F})\gg a$,
with $a$ the s-wave
scattering length, we can use a pseudo-potential \cite{AshcroftMermin} 
for a single electron-impurity interaction. In $d=3$,
\begin{equation}
v({\bf x}-{\bf R}_{i}) = \frac{4\pi a}{m}\,\delta ({\bf x}-{\bf R}_{i})\quad. 
\label{eq:2.2c}
\end{equation}
%\end{mathletters}%
The relevant scattering potential is given by
$u({\bf x})=v({\bf x}) - \{v({\bf x})\}_{\rm dis}$. Since the mean,
$\{v({\bf x})\}_{\rm dis}$, simply redefines the Fermi energy, we can ignore
it and have
%\begin{equation}
$$
\qquad\qquad\qquad\qquad\left\{ u({\bf x})u({\bf y})\right\}_{\rm dis} =
    \left(\frac{4\pi }{m}\right)\,a^{2}\,n_{i}\,\delta ({\bf x-y})
                        \quad.\qquad\qquad\quad\ (2')
$$
%\end{equation}
Here $n_{i}=N/V$ is the impurity density. Using $\tau\sim (n_{i}\,a^{2})^{-1}$, 
in $d=3$, we see that Eq.\ (2') has the same structure as 
Eq.\ (\ref{eq:2.1}). The reason for the other factors in 
Eq.\ (\ref{eq:2.1}) will become clear below.

Finally, we write down the Hamiltonian for the Edwards model in
second quantization \cite{AGD},
\begin{equation}
\hat{H}=\int d{\bf x}\,\hat{\psi }_{\sigma}^{\dagger}({\bf x})
   \left( -\frac{\nabla ^{2}}{2m}\right) \hat{\psi }_{\sigma }({\bf x})
     + \int d{\bf x}\,u({\bf x})\,\hat{\psi}_{\sigma }^{\dagger}({\bf x})\,
         \hat{\psi}_{\sigma}({\bf x})\quad.
\label{eq:2.3}
\end{equation}
Here $\hat{\psi}_{\sigma}^{\dagger}$ and $\hat{\psi}_{\sigma}$ are
fermion creation and annihilation operators
with $\sigma$ denoting a spin label,
and summation over repeated Greek labels is understood.

\subsection{Linear density response}
\label{subsec:II.B}

A standard technique in nonequilibrium statistical mechanics is to apply an
external field to equilibrium systems, and to calculate the response of the
system to this field \cite{FetterWalecka}. 
By expanding in powers of the external field, nonequilibrium
quantities can be related to equilibrium time correlation functions. For
example, let us apply an electric field ${\bf E}$ to an electron gas and
expand to linear order in the field. Using Ohm's law,
${\bf J} = \sigma \cdot {\bf E}$, which relates the induced current ${\bf J}$
to the field, we obtain the so-called Kubo formula for the frequency
dependent conductivity \cite{Kubo},
\begin{equation}
\sigma (\omega) = -\frac{n_{e}\,e^{2}}{m\,i\omega} + \frac{1}{\omega V}
\int\limits_{0}^{\infty }dt\ e^{i\omega t}\,\left\{\left\langle 0\left\vert
   \left[{\hat J}_{x}(t),{\hat J}_{x}(t=0)\right]\,\right\vert 0\right\rangle 
                    \right\}_{\rm dis}\quad. 
\label{eq:2.4}
\end{equation}
Here $n_{e}$ is the electron density, ${\hat J}_{x}(t)$ is the
total current operator at time $t$ in the $x$-direction,
$[{\hat a},{\hat b}]$ denotes the commutator of operators
${\hat a}$ and ${\hat b}$,
and $\left\langle 0\left\vert\ldots\right\vert 0\right\rangle$ 
denotes a ground state quantum mechanical expectation value.

Similarly, if we apply a potential $\varphi ({\bf x},t)$ that couples to
the density, then the change in the electronic density, $\delta n$, is given
by,
%\begin{mathletters}
\label{eqs:2.5}
\begin{equation}
\delta n({\bf k},\omega) = \chi_{nn}(k,\omega)\,\varphi ({\bf k},\omega)\quad,
\label{eq:2.5a}
\end{equation}
with $\chi_{nn}$ the density response function and $k = \vert{\bf k}\vert$.
$\chi_{nn}(k,\omega)$ depends on $k$ rather than on ${\bf k}$ because of
rotational invariance once the disorder average has been performed. It
is the Fourier transform of the density-density correlation function
\begin{equation}
\chi_{nn}({\bf x},t) = \Theta (t)\ \biggl\{\Bigl\langle 0\Bigl\vert\bigl[
   {\hat n}({\bf x}=0,t=0),{\hat n}({\bf x},t)\bigr]\Bigr\vert 0\Bigr\rangle
        \biggr\}_{\rm dis}\quad .  
\label{eq:2.5b}
\end{equation}
%\end{mathletters}%
Here $\hat{n}$ is the density operator.

Equations\ (\ref{eq:2.4}) and (\ref{eq:2.5b}) are valid for 
arbitrary systems. For
noninteracting systems, some simplifications apply. In particular, the
remaining many-body aspect of the problem (i.e., the correlations due
to the Pauli principle) can be factored out, and the problem can be reduced to
one of a {\em single} particle moving in a random potential. 
Technically, this is
possible because the field operators ($\hat{\psi}^{\dagger}$ and $\hat{\psi}$)
can be expanded in an exact eigenstate basis in terms of creation and
annihilation operators ${\hat a}^{\dagger}$ and ${\hat a}$. The net result is
that the quantum mechanical averages in Eqs.\ (\ref{eq:2.4}), (\ref{eq:2.5b})
factorize. For example, the real part of the conductivity can be written
%\begin{mathletters}
\label{eqs:2.6}
\begin{eqnarray}
\sigma'(\omega) \equiv {\rm Re}\sigma (\omega)&=&\frac{e^{2}}{2\pi Vm^{2}}
{\rm Re}\int d{\bf p}\,d{\bf p}_{1}\,p_{x}\,p_{1x}\,
         \left[\left\{ G^{R}({\bf p,p}_{1},E_{\rm F}+\omega)\,\right.\right.
\nonumber\\
    &&\qquad\qquad\qquad\qquad\qquad\qquad\times\left. 
          G^{A}({\bf p}_{1},{\bf p},E_{\rm F})\right\}_{\rm dis}
\nonumber\\
     && - \left.\left\{ G^{A}({\bf p,p}_{1},E_{\rm F}+\omega)\,
       G^{A}({\bf p}_{1},{\bf p},E_{\rm F})\right\}_{\rm dis} \right] \quad,
\label{eq:2.6a}
\end{eqnarray}
with
\begin{equation}
G^{R,A}({\bf p}_{1},{\bf p},E) = \left\langle{\bf p}\left\vert\,
    \frac{1}{E-H\pm i0}\,\right\vert {\bf p}_{1}\right\rangle\quad,
\label{eq:2.6b}
\end{equation}
denoting retarded (R) and advanced (A) Green functions. 
$\vert{\bf p}\rangle$ denotes a single-particle plane wave state, and 
$\hat H$ is the single-particle Hamiltonian,
\begin{equation}
{\hat H} = -\frac{\nabla^{2}}{2m} + u({\bf x})\quad.  
\label{eq:2.6c}
\end{equation}
%\end{mathletters}%
Similarly, the density response function can be written,
\begin{eqnarray}
\chi_{nn}(k,\omega) = N_{\rm F}&+&\frac{i\omega}{2\pi}\int d{\bf p}\,
     d{\bf p}_{1} \,
\left\{ G^{R}({\bf p+k},{\bf p}_{1}+{\bf k},E_{\rm F}
   + \omega)\,\right.
\nonumber\\
&&\qquad\qquad\qquad\qquad\times\left. G^{A}({\bf p}_{1}, {\bf p},E_{\rm F})
                    \right\}_{\rm dis}\quad.
\label{eq:2.7}
\end{eqnarray}

\subsection{Some additional remarks}
\label{subsec:II.C}

In the long-wavelength limit, the form of $\chi_{nn}(k,\omega)$ is
determined by the diffusive nature of the electron dynamics in 
a disordered solid. This can be seens as follows. The conservation 
law for the particle number density
is expressed by the continuity equation
%\begin{mathletters}
\label{eqs:2.8}
\begin{equation}
\partial_t\,\delta n({\bf x},t) + {\bf\nabla}\cdot{\bf j}({\bf x},t) = 0\quad,
\label{eq:2.8a}
\end{equation}
with ${\bf j}$ the number current density. In addition, Fick's law, valid
in the long-wavelength limit, gives a relation between ${\bf j}$,
$\delta n$, and $\varphi$,
\begin{equation}
{\bf j}({\bf x},t) = -D\,{\bf\nabla}\delta n + \mu_{\rm e}{\bf\nabla}\varphi 
                        \quad.
\label{eq:2.8b}
\end{equation}
%\end{mathletters}%
with $D$ the diffusion coefficient and 
$\mu_{\rm e} =D\partial n/\partial\mu =DN_{\rm F}$ the electron mobility
(not to be confused with the chemical potential $\mu$). 
Using Eq.\ (\ref{eq:2.8b}) in Eq.\ (\ref{eq:2.8a}), performing a
Fourier-Laplace transform, and comparing with Eq.\ (\ref{eq:2.5a}) gives
\begin{equation}
\chi_{nn}(k,\omega) = \frac{N_{\rm F}Dk^2}{-i\omega +Dk^2}\quad.
\label{eq:2.9}
\end{equation}

The above arguments tell us the functional form of $\chi_{nn}$ in the
long-wavelength small-frequency limit, but they do not tell us the value
of the diffusivity $D$. Rather, $D$ enters as a purely phenomenological
quantity. However, we can get a rough idea of the value of $D$, and of the
closely related conductivity, by means of simple dimensional analysis.
The units of $D$ are $L^{2}/T$, with $L$ a length scale and $T$ a
time scale. There are three length scales in the problem: The Fermi
wavelength $\lambda_{\rm F}$, the scattering length $a$, and the
scattering mean-free path $\ell$. Clearly, the relevant length for
transport processes is $\ell$. Similarly, the relevant time scale is
the scattering mean-free time $\tau$. For electrons on the Fermi surface, 
$\ell = v_{\rm F}\tau$, with $v_{\rm F} = k_{\rm F}/m$ the Fermi velocity.
As an order of magnitude estimate of $D$ we therefore expect
%\begin{mathletters}
\label{eqs:2.10}
\begin{equation}
D \simeq v_{\rm F}^{2}\tau \quad .  
\label{eq:2.10a}
\end{equation}
Using Eq.\ (\ref{eq:1.1}) and $v_{\rm F}^{2}N_{\rm F}\simeq n_{e}/m$, 
the analogous
estimate for the conductivity $\sigma$ is
\begin{equation}
\sigma \simeq \frac{n_{e}e^{2}\tau}{m}\quad .  
\label{eq:2.10b}
\end{equation}
%\end{mathletters}%
This is the standard Drude result for $\sigma$.

Clearly, all of the above arguments and derivations are just phenomenological
in nature, and therefore ultimately unsatisfactory, although they do contain
the correct physics. In the following section we improve our treatment, and
derive the Eqs.\ (\ref{eq:2.9}), (\ref{eq:2.10a}), and (\ref{eq:2.10b})
from the microscopic Hamiltonian.

\vskip 3mm
\section{Many Body Perturbation Theory}
\label{sec:III}
\vskip 1mm
\subsection{The disorder averaged Green function}
\label{subsec:III.A}

For pedagogical reasons, let us first calculate the averaged Green function,
although it is not directly needed to compute either 
Eq.\ (\ref{eq:2.6a}) or Eq.\ (\ref{eq:2.7}).
Since on average space is homogeneous, we have
\begin{equation}
\left\{ G^{R,A}({\bf p},{\bf p}_{1},E)\right\}_{\rm dis}
    = \delta ({\bf p}-{\bf p}_{1})\, G^{R,A}({\bf p},E)\quad .  
\label{eq:3.1}
\end{equation}
A self-energy $\Sigma$ can be defined in the usual way, leading
to a Dyson equation for the Green function \cite{FetterWalecka}. 
To be specific, we consider the retarded Green function,
\begin{eqnarray}
G^{R}({\bf p},E) &=& G_{0}^{R}({\bf p},E) + G_{0}^{R}({\bf p},E)\,
                    \Sigma^{R}({\bf p},E)\,G^{R}({\bf p},E) 
\nonumber\\
                 &=&\frac{1}{E - {\bf p}^{2}/2m - \Sigma^{R}({\bf p},E) + i0}
                                                                  \quad ,  
\label{eq:3.2}
\end{eqnarray}
with $G_{0}$ the free particle Green function (i.e., Eq.\ (\ref{eq:3.2})
with $\Sigma^R = 0$).

\begin{figure}[ht]
\vskip 10mm
\centerline{\psfig{figure=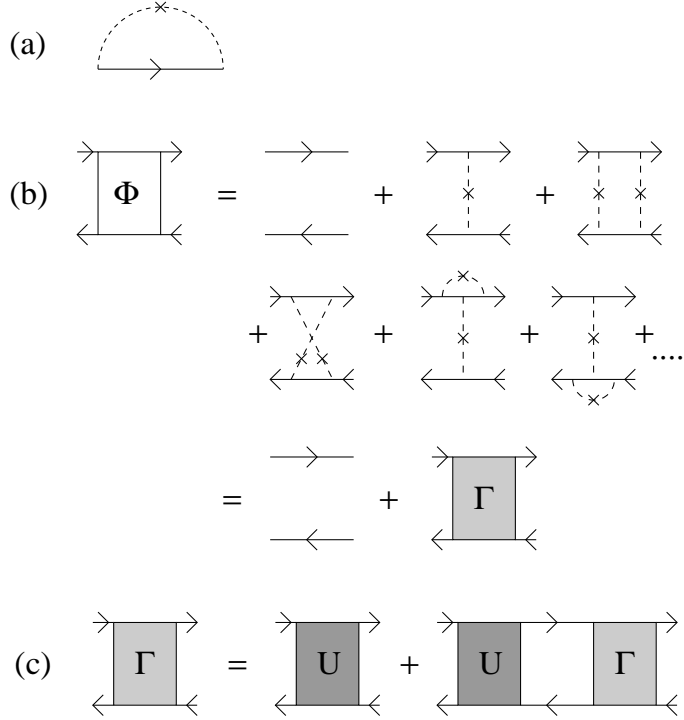,width=90mm}\vspace*{5mm}}
\vskip -5mm
\caption{(a) Born approximation for the electron self-energy. (b) Perturbation
theory for the two-particle correlation function $\Phi$, 
Eq.\ (\protect\ref{eq:3.4}). (c) The reducible vertex function $\Gamma$
in terms of the irreducible vertex function $U$.}
\label{fig:1}
\end{figure}
Standard methods lead to a diagrammatic theory for $\Sigma^R$ \cite{Edwards}. 
In Fig.\ \ref{fig:1} (a), we show the lowest order, or Born, 
approximation for the self-energy. Here the straight
line denotes a free particle Green function and the dashed line with a cross
denotes a $\{u\, u\}_{\rm dis}$ disorder correlation function. 
In this approximation,
%\begin{mathletters}
\label{eqs:3.3}
\begin{eqnarray}
\Sigma^{R}({\bf p},E) = \frac{1}{2\pi N_{\rm F}\tau}\int\limits_{\bf q}
                             \frac{1}{\left[E - {q^{2}}/{2m}+i0\right]} 
  &\approx&\frac{-i\pi}{2\pi N_{\rm F}\tau }\int\limits_{\bf q}\delta\left( 
          E-\frac{q^{2}}{2m}\right) 
\nonumber\\
  &=&-\frac{i}{2\tau }\quad ,  
\label{eq:3.3a}
\end{eqnarray}
Here $\int\limits_{\bf q} \equiv \int d{\bf q}/(2\pi)^d$, and
we have dropped the real part of $\Sigma^R$, which simply
redefines the zero of the energy. In this approximation the average Green
function is,
\begin{equation}
G^{R}({\bf p},E)\approx \frac{1}{E - {\bf p}^{2}/2m + i/2\tau}\quad . 
\label{eq:3.3b}
\end{equation}
%\end{mathletters}%
We see that the factors in Eq.\ (\ref{eq:2.1}) 
were chosen so that the time Fourier transform of Eq.\ (\ref{eq:3.3b}),
$G^{R}({\bf p},t)$, decays exponentially with a relaxation time equal to 
$\tau$. The physical meaning of this decay is the loss of phase correlations
that is brought about by the electron-impurity scattering.

\subsection{Theory for the conductivity and the density susceptibility: 
            Generalized Boltzmann equation}
\label{subsec:III.B}

In order to calculate the conductivity and the density susceptibility,
Eqs.\ (\ref{eq:2.6a}) and (\ref{eq:2.7}), we introduce the 
two-particle correlation function,
\begin{equation}
\Phi_{{\bf pp}_{1}}({\bf k},\omega) = \left\{ G^{R}\left( {\bf p}+{\bf k},
{\bf p}_{1}+{\bf k},E_{\rm F}+\omega \right)\, G^{A}\left( {\bf p}_{1},{\bf p}
,E_{\rm F}\right) \right\}_{\rm dis} \quad .  
\label{eq:3.4}
\end{equation}
The advanced-advanced combination $G^A\,G^A$ also contributes to
Eq.\ (\ref{eq:2.6a}), but it is not important for the effects we are
intererested in and we therefore do not dicuss it.

A diagrammatic expression for $\Phi$ is shown in 
Fig.\ \ref{fig:1} (b). Here the straight
lines denote exact averaged Green functions, and the shaded box indicates
disorder correlations between the top (retarded) and bottom (advanced) Green
functions. A Bethe-Salpeter equation can be used to introduce a 
two-particle irreducible
vertex, $U_{{\bf pp}_{1}}$, which is the two-particle analog of the Dyson
self-energy. The diagrammatic relationship between the reducible vertex
$\Gamma$, and the irreducible one is shown in Fig.\ \ref{fig:1} (c). 
Analytically we have
\begin{eqnarray}
\Phi_{{\bf pp}_{1}}({\bf k},\omega)&=&G^{R}\left( {\bf p}+{\bf k}
,E_{\rm F}+\omega \right)\,G^{A}\left({\bf p},E_{\rm F}\right)
\biggl[\delta ({\bf p-p}_{1})
\nonumber\\
&&\qquad\qquad\qquad\qquad + \int\limits_{{\bf p}_{2}}
   U_{{\bf pp}_{2}}({\bf k},\omega )\Phi_{{\bf p}_{2}{\bf p}_{1}}
                                ({\bf k},\omega )\biggr] \ .  
\label{eq:3.5}
\end{eqnarray}
To put this into a more standard form we write \cite{VW},
%\begin{mathletters}
\label{eqs:3.6}
\begin{equation}
G^{R}\left( {\bf p}+{\bf k},E_{\rm F}+\omega\right)\, 
    G^{A}\left( {\bf p},E_{rm F}\right) 
= -\frac{\Delta G_{{\bf p}}({\bf k},\omega )}{\omega -
      {\bf k\cdot p}/{m}-\Delta \Sigma _{{\bf p}}({\bf k},\omega )}\quad , 
\label{eq:3.6a}
\end{equation}
where
\begin{equation}
\Delta G_{{\bf p}}({\bf k},\omega )=G^{R}\left( {\bf p}+{\bf k},E_{\rm F}
+\omega \right) -G^{A}\left( {\bf p},E_{\rm F}\right)  \quad,
\label{eq:3.6b}
\end{equation}
\begin{equation}
\Delta \Sigma_{{\bf p}}({\bf k},\omega) = \Sigma^{R}({\bf p}+{\bf k}
,E_{\rm F}+\omega )-\Sigma^{A}({\bf p},E_{\rm F})\quad .  
\label{eq:3.6c}
\end{equation}
%\end{mathletters}%
Defining $\widetilde{\Phi}$ through
%\begin{mathletters}
\label{eqs:3.7}
\begin{equation}
\Phi_{{\bf pp}_{1}}({\bf k},\omega) = \Delta G_{{\bf p}}({\bf k},\omega)\,
\widetilde{\Phi}_{{\bf pp}_{1}}({\bf k},\omega) \quad, 
\label{eq:3.7a}
\end{equation}
and using the Ward identity \cite{VW},
\begin{equation}
\Delta \Sigma _{{\bf p}}({\bf k},\omega )=\int\limits_{{\bf p}_{2}}U_{{\bf pp
}_{2}}({\bf k},\omega )\,\Delta G_{{\bf p}_{2}}({\bf k},\omega )\quad , 
\label{eq:3.7b}
\end{equation}
%\end{mathletters}%
the Eq.\ (\ref{eq:3.5}) can be written,
\begin{eqnarray}
\left\{ \omega -\frac{{\bf k\cdot p}}{m}\right\} \widetilde{\Phi }_{{\bf pp}
_{1}}({\bf k},\omega) &=& -\delta ({\bf p-p}_{1}) 
+\int\limits_{{\bf p}_{2}}U_{{\bf pp}_{2}}({\bf k},\omega)\,\Delta G_{{\bf p}
_{2}}({\bf k},\omega)
\nonumber\\
&&\qquad\times\left[ \widetilde{\Phi }_{{\bf pp}_{1}}({\bf k},\omega
)-\widetilde{\Phi }_{{\bf p}_{2}{\bf p}_{1}}({\bf k},\omega )\right] \quad . 
\label{eq:3.8}
\end{eqnarray}

Two comments are in order. First, for small wavevectors and frequencies, 
$$\Delta G_{\bf p}({\bf k}\rightarrow 0,\omega\rightarrow 0)\approx 
   -i\pi \delta (E-p^{2}/2m)/\tau\quad,$$ 
so that the $\Delta G$ in Eqs.\ (\ref{eq:3.8}) and (\ref{eq:3.6a}) 
are delta-functions that confine the
electrons to the energy shell. Second, Eq.\ (\ref{eq:3.8}) for $\Phi$ is exact,
and has the form of a generalized Boltzmann
equation. That is, the frequency term on the left hand side is the Fourier
transform of a time derivative, and ${\bf k}\cdot {\bf p}/m$ is the spatial
Fourier transform of ${\bf p}\cdot {\bf \nabla }/m$. 
Therefore the left-hand side has the form of a standard
free-streaming term. The right-hand side, on the other hand, is an initial
condition term, plus a generalized collision operator. In the next
subsection, we will see that the simplest approximation for $U$ leads to
the quantum Boltzmann equation.

In terms of $\widetilde\Phi$, the physical quantities of interest to us are
%\begin{mathletters}
\label{eqs:3.9}
\begin{equation}
\sigma'(\omega) = \frac{e^{2}}{2\pi Vm^{2}}\int d{\bf p}\,d{\bf p}_{1}\,
         p_{x}\,p_{1x}\,\Delta G_{{\bf p}}({\bf 0},\omega )\,
  \widetilde{\Phi }_{{\bf pp}_{1}}({\bf 0},\omega )\quad ,  
\label{eq:3.9a}
\end{equation}
and
\begin{equation}
\chi_{nn}(k,\omega) = N_{\rm F} + \frac{i\omega}{2\pi}\int d{\bf p}\,
   d{\bf p}_{1}\,\Delta G_{{\bf p}}({\bf k},\omega)\,
       \widetilde{\Phi}_{{\bf pp}_{1}}({\bf k},\omega)\quad .  
\label{eq:3.9b}
\end{equation}
%\end{mathletters}%

\subsection{Quantum Boltzmann equation}
\label{subsec:III.C}

\begin{figure}[ht]
\centerline{\psfig{figure=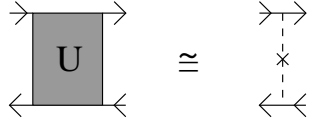,width=40mm}\vspace*{5mm}}
\vskip -5mm
\caption{Boltzmann approximation for the irreducible vertex function $U$.}
\label{fig:2}
\end{figure}

To lowest order in the disorder we have, cf. Fig.\ \ref{fig:2},
\begin{equation}
U_{{\bf pp}_{2}}({\bf k},\omega )=\frac{i}{2\pi N_{\rm F}\tau }\quad.  
\label{eq:3.10}
\end{equation}
Inserting Eqs.\ (\ref{eq:3.10}) in Eq.\ (\ref{eq:3.8}) yields,
\begin{eqnarray}
\left\{ -i\omega +i\frac{{\bf k\cdot p}}{m}\right\} \widetilde{\Phi }_{{\bf 
pp}_{1}}({\bf k},\omega) &=& i\delta ({\bf p-p}_{1}) 
+\frac{1}{N_{\rm F}\tau }\int\limits_{{\bf p}_{2}}\delta
(E_{\rm F}-p_{2}^{2}/2m)
\nonumber\\
&&\quad\times\left[ \widetilde{\Phi }_{{\bf p}_{2}{\bf p}_{1}}({\bf k}
,\omega )-\widetilde{\Phi }_{{\bf pp}_{1}}({\bf k},\omega )\right]\,. 
\label{eq:3.11}
\end{eqnarray}
This is the quantum Boltzmann equation, which thus turns out to be
the simplest nontrivial approximation within the formalism developed above.
Higher order approximations for $U_{{\bf pp}_{2}}$ lead
to corrections of order $\lambda_{\rm F}/\ell $ and $a/\ell$ 
to the Boltzmann equation results.

The Eq.\ (\ref{eq:3.11}) can be easily solved exactly. One obtains,
%\begin{mathletters}
\label{eqs:3.12}
\begin{eqnarray}
\widetilde{\Phi }_{{\bf pp}_{1}}({\bf k},\omega) &=&
   i\delta ({\bf p-p}_{1})\,h_{{\bf p}}({\bf k},\omega)
\nonumber\\ 
 && + \frac{i\,\delta (E_{\rm F}-p_{1}^{2}/2m)\,h_{{\bf p}}({\bf k},\omega)}
    {N_{\rm F}\tau\left[ 1-\frac{1}{N_{\rm F}\tau}  
        \int\limits_{{\bf p}_{2}}\delta              
(E_{\rm F}-p_{2}^{2}/2m)\,h_{{\bf p}_{2}}({\bf k},\omega )\right] }\quad , 
\label{eq:3.12a}
\end{eqnarray}
with
\begin{equation}
h_{{\bf p}_{2}}({\bf k},\omega )=\left( -i\omega +i\,\frac{{\bf k\cdot p}}{m}+
\frac{1}{\tau }\right)^{-1}\quad .  
\label{eq:3.12b}
\end{equation}
%\end{mathletters}%
For general ${\bf k}$ and $\omega$, Eq.\ (\ref{eq:3.12a}) is quite
complicated. However, using it in Eqs.\ (\ref{eq:3.9a}) and (\ref{eq:3.9b}) 
for small wavenumber and frequency immediately gives
%\begin{mathletters}
\label{eqs:3.13}
\begin{equation}
\sigma_{\rm B}=n_{e}\frac{e^{2}\tau }{m} \quad, 
\label{eq:3.13a}
\end{equation}
and,
\begin{equation}
\chi _{nn,{\rm B}}({\bf k},\omega )\simeq 
    \frac{N_{\rm F}D_{\rm B}k^{2}}{-i\omega
+D_{\rm B}k^{2}}\quad ,  
\label{eq:3.13b}
\end{equation}
where the subscript B denotes the Boltzmann approximation. Notice that
$\sigma_{\rm B}$ and $D_{\rm B}$ are related by the Einstein relation,
\begin{equation}
\sigma_{\rm B}=e^{2}N_{\rm F}D_{\rm B}\quad ,
\label{eq:3.13c}
\end{equation}
%\end{mathletters}%
as they should be according to Eq.\ (\ref{eq:1.1}).

\section{Corrections to Boltzmann Transport Theory: Wave Interference Effects}
\label{sec:IV}

\subsection{Disorder or density expansion of $\sigma$}
\label{subsec:IV.A}

Perhaps the most obvious theoretical problem concerning the wave
interference effects is to characterize the corrections to
$\sigma_B$ that arise when $\lambda_{\rm F}/\ell$ is not zero.
The first question is whether or not there is an analytic
expansion in the small parameter $\epsilon =\lambda_{\rm F}/\ell$.
Technically, this question can be answered by examining the coefficients 
$a_i\ (i=1,2,3,\ldots)$ in the Taylor expansion,
\begin{equation}
\frac{\sigma}{\sigma_{\rm B}} = 1 + a_{1}\,\epsilon + a_{2}\,\epsilon^2 
    +O(\epsilon^3)\quad . 
\label{eq:4.1}
\end{equation}
Detailed calculations show that in general such an expansion
does not exist, and that all the coefficients beyond $a_{d-1}$ are infinite
in $d$ dimensions \cite{density_expansion}. 
Although of different physical origin, a similar divergence
is well known to exist in the kinetic theory of classical 
gases \cite{DKS}.

\begin{figure}[ht]
\centerline{\psfig{figure=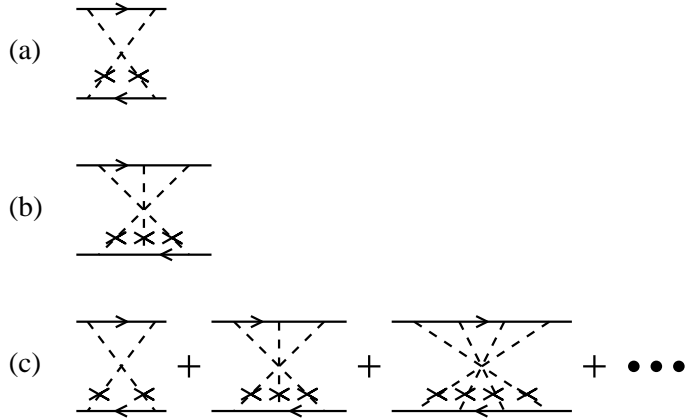,width=90mm}\vspace*{5mm}}
\vskip -5mm
\caption{Maximally crossed diagrams.}
\label{fig:3}
\end{figure}
To see the origin of these infinities in the quantum case, consider the
first few maximally crossed diagrams (MCD) in Fig.\ \ref{fig:3}. 
The calculation gives,
%\begin{mathletters}
\label{eqs:4.2}
\begin{equation}
(3a) \propto \int\limits_{q}G^{R}({\bf q})\,G^{A}({\bf p}+{\bf p}_{1}-{\bf q}) 
\biggl\vert_{\omega=1/\tau=0}
\propto \frac{1}{\left\vert {\bf p}+{\bf p}_{1}\right\vert }  
\label{eq:4.2a}
\end{equation}

\begin{equation}
(3b)\propto (3a)^{2}\propto\frac{1}{\left\vert{\bf p}+{\bf p}_{1}\right\vert
                                  ^{2}}\quad.
\label{eq:4.2b}
\end{equation}
%\end{mathletters}%
That is, the MCD are singular for the momentum configuration that
corresponds to
backscattering, ${\bf p}_{1} = -{\bf p}$. In any dimension, 
these singularities
lead to infinite coefficients in Eq.\ (\ref{eq:4.1}), with the order
in which the first infinity occurs depending on the dimension.
The correct expansion of $\sigma$ in powers of $\epsilon$ is
therefore nonanalytic. In $d=3$, it is of the form
\begin{equation}
\frac{\sigma }{\sigma_{\rm B}} = 1 + a_{1}\,\epsilon +a_{2}^{\prime}\,
   \epsilon^{2}\ln 1/\epsilon + a_{2}\,\epsilon^{2} + \ldots \quad. 
\label{eq:4.3}
\end{equation}
In $d=2$ it has been shown that power-law nonanalyticities appear in
addition to logarithmic ones \cite{2d_density_expansion}.
In the next subsection we further discuss the backscattering 
processes that cause these nonanalyticities.

\begin{figure}[ht]
\centerline{\psfig{figure=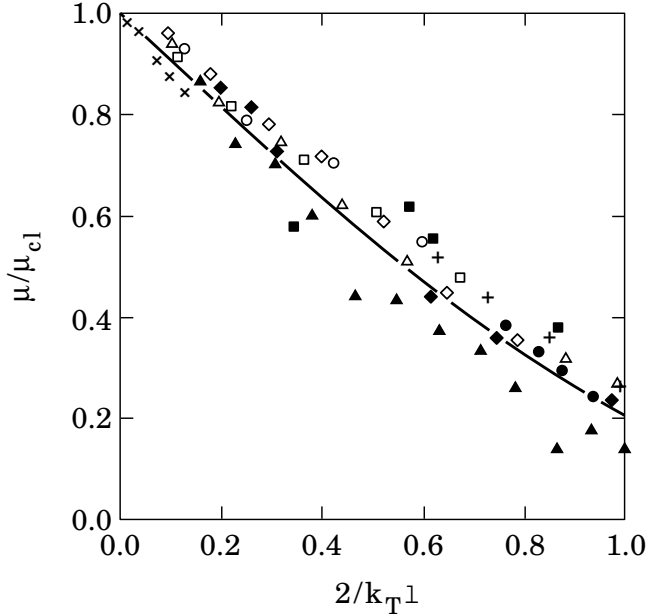,width=90mm}\vspace*{5mm}}
\vskip -5mm
\caption{Mobility $\mu$ of electrons in dense gases, normalized to the
 Boltzmann value $\mu_{\rm cl}$, as a function of $\chi$. The symbols
 represent experimental data as measured or quoted by Adams at al. 
 \protect\cite{Adams}. The solid line is the theoretical result,
 Eqs.\ (\protect\ref{eq:4.5},\protect\ref{eq:4.6}). From
 Ref.\ \protect\cite{us_ernst}.}
\label{fig:4}
\end{figure}

For most systems, the values of the coefficients in Eq.\ (\ref{eq:4.3}) are of
limited interest because the derivation has ignored
electron-electron interactions, which are important in
all solid state materials. However, there does exist a physical system for which
the quantum Lorentz gas is a realistic model, namely electrons injected
into He gas at low ($\leq 4{{}^{\circ }}{\rm K})$ temperatures.
In Ref.\ \cite{us_PRE} detailed arguments have been given
that on electronic time scales, the He
atoms can be treated as stationary scatterers and that all other
approximations used in our model are justified for describing transport in
this system. The finite temperature electron mobility, $\mu$, normalized to
its Boltzmann value, can be expanded in the small parameter
\begin{equation}
\chi = \lambda _{T}/\pi \ell \quad ,  
\label{eq:4.4}
\end{equation}
with $\lambda_T\equiv 2\pi/k_T = \sqrt{2\pi^2/mT}$ 
the thermal de Broglie wavelength. The result is,
\begin{equation}
\frac{\mu}{\mu_{\rm B}} = 1 + \mu_{1}\,\chi + \mu_{2}^{\prime}\,\chi^{2}\ln\chi
   +\mu_{2}\,\chi^{2} + \ldots \quad , 
\label{eq:4.5}
\end{equation}
with,
\begin{eqnarray}
\mu_{1} &=& - \pi^{3/2}/6 \quad,
\nonumber\\
\mu_{2}^{\prime} &=& (\pi^{2}-4)/32\quad,
\label{eq:4.6}\\
\mu _{2} &=& 0.236\quad . 
\nonumber
\end{eqnarray}
In Fig.\ \ref{fig:4} this theoretical result, with no adjustable parameters,
is compared to experimental data. We conclude that experiments are consistent
with Eqs.\ (\ref{eq:4.5}) and (\ref{eq:4.6}). Further
experiments are needed to unambigiously confirm the 
presence of the logarithm in these
equations. For further discussions of this point, see 
Ref.\ \cite{us_PRE}.

\subsection{Weak localization effects and long-time tails}
\label{subsec:IV.B}

In classical transport theory it has proven to be very enlightening to
examine the long-time properties of the time correlation functions (TCF) that
determine the transport coefficients. Generally, it has been found that
these functions decay algebraically for long times and that therefore the
transport coefficients themselves are nonanalytic functions of the
frequency \cite{DKS}. 
In quantum transport theory the decay of the TCF has proven to be
even more interesting \cite{us_ernst}, 
and to be closely related to the phenomenon of
Anderson localization \cite{LeeRama}. 
Note that in either case, these algebraic tails are
qualitatively different from the exponential decays predicted by both the
Boltzmann equation and by simple theories for the Green function.

Here we mainly concentrate on a physical argument \cite{Bergmann}
that leads to these
so-called weak localization effects. Some hints of their technical origin
will also be given; details can be found elsewhere \cite{LeeRama}.
We will see that
the weak localization effects are closely related to the same backscattering 
phenomena that are responsible
for the nonanalytic disorder expansion discussed in the previous subsection.

\begin{figure}[ht]
\vskip 30mm
\centerline{\psfig{figure=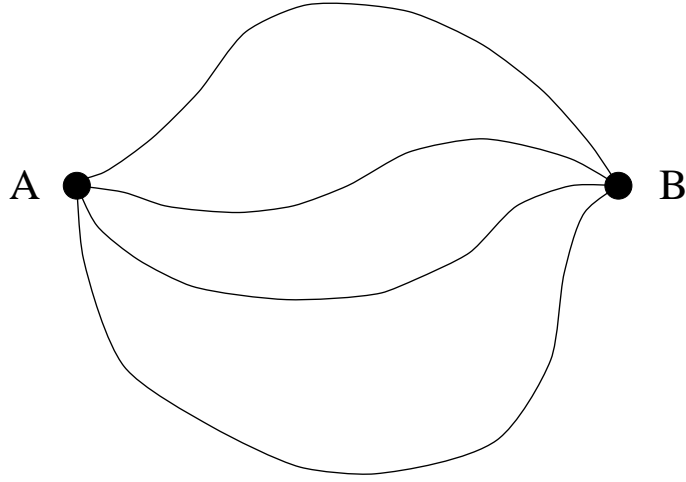,width=90mm}\vspace*{5mm}}
\vskip -25mm
\caption{Feynman paths from point $A$ to point $B$.}
\label{fig:5}
\end{figure}
The physical argument starts by considering the different Feynman paths
from, say, point A to point B, as illustrated in Fig.\ \ref{fig:5}.
Labeling the
amplitude of path $i$ by ${\cal A}_{i}$, 
the total probability to reach B from A is,
\begin{equation}
w = \left\vert\sum_{i} {\cal A}_{i}\right\vert^{2}
  = \sum_{i}\left\vert {\cal A}_{i}\right\vert^{2}
       + \sum_{i\neq j}{\cal A}_{i}\,{\cal A}_{j}^{*}\quad .  
\label{eq:4.7}
\end{equation}
Usually, the different paths are uncorrelated and the second term
in Eq.\ (\ref{eq:4.7}) averages to zero, 
leaving the first term, which is just the
classical probability. The exception to this is if the points A and B
coincide. In this case the path can be traversed in two opposite directions,
namely, forward or backward. The amplitudes 
${\cal A}_{1}$ and ${\cal A}_{2}$ for these
two paths have a coherent phase relation, leading to constructive
interference. If ${\cal A}_{1} = {\cal A}_{2} = {\cal A}$, 
then Eq.\ (\ref{eq:4.7}), for $i=1,2$, becomes
%\begin{equation}
$$
\qquad\qquad\qquad w = 2\left\vert {\cal A}\right\vert^{2} + 
        2{\cal A}\,{\cal A}^{*}
  = 4\left\vert {\cal A}\right\vert^{2} \quad,\qquad\qquad\qquad
                               \qquad\qquad\quad (47')
$$
%\end{equation}
which is twice the
classical probability. The important conclusion from this argument is that
the quantum probabilities for 
time reversed, or intersecting, paths are enhanced compared to classical
dynamics. Further, since paths that return to their starting point slow down
the diffusion process, we expect these quantum corrections to lead to a
decrease of $\sigma$ compared to the semi-classical value.

To estimate the contribution of the closed loops to $\sigma$ we argue as
follows. If the particle is diffusing, then the probability of finding it 
at point
${\bf r}$ at time $t$, given that it was at point 
${\bf r}_{0}$ at time $t=0$ is
%\begin{mathletters}
\label{eqs:4.8}
\begin{equation}
w({\bf r},t) = \frac{\exp [-({\bf r}-{\bf r}_{0})^{2}/4Dt]}{(4\pi Dt)^{d/2}}
              \quad,  
\label{eq:4.8a}
\end{equation}
$w({\bf r},t)$ is appreciably different from zero only within a diffusive 
volume $V_{\rm diff}$ determined by $({\bf r}-{\bf r}_{0})^{2}<4Dt,$ or,
\begin{equation}
V_{\rm diff}\approx (Dt)^{d/2}\quad .  
\label{eq:4.8b}
\end{equation}
%\end{mathletters}%
\begin{figure}[ht]
\centerline{\psfig{figure=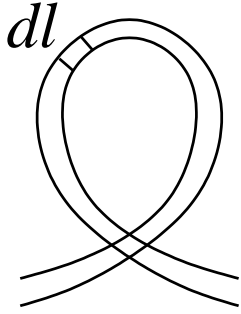,width=50mm}\vspace*{5mm}}
\vskip -5mm
\caption{Geometry of a Feynman tube.}
\label{fig:6}
\end{figure}
Now, each Feynman path has a diameter proportional to $\lambda_{\rm F}^{d-1}$,
and a differential length $d\ell = v_{\rm F}\,dt$, cf. Fig.\ \ref{fig:6}.
The differential
volume of the Feynman tube is $dV=V_{\rm F}\lambda _{p}^{d-2}dt,$ which should
be compared to $V_{\rm diff}$. If $p$ is the probability that a closed path
exists, then it is the ratio of these two volumes integrated over all times:
\begin{equation}
p = \int\limits_{\tau ^{\prime }}^{T}\frac{dV}{V_{\rm diff}}
  =\lambda_{\rm F}^{d-1}v_{\rm F}\int\limits_{\tau ^{\prime }}^{T}
          \frac{dt}{(Dt)^{d/2}}\quad .
\label{eq:4.9}
\end{equation}
Here $\tau^{\prime } = O(\tau)$ is a microscopic time where the
diffusive description breaks down, and $T$ is the observation time that we
will equate with $w^{-1}$. The change in $\delta\sigma$ of the conductivity
due to these wave interference effects is therefore,
\begin{equation}
\frac{\delta\sigma}{\sigma_0} \propto - \cases{
   1/(\omega\tau)^{1/2}\quad & ,\quad$(d=1)$\cr
   \ln (1/\omega\tau)\quad    & ,\quad$(d=2)$\cr
   {\rm const.} - (\omega\tau)^{1/2}\quad  &  ,\quad$(d=3)$\cr}\quad.
%\frac{\delta \sigma }{\sigma _{0}}\sim -\left\{ 
%\begin{tabular}{l}
%$\frac{1}{(\omega \tau )^{1/2}}\quad \quad d=1\vspace*{0.1in}$ \\ 
%\ln$\frac{1}{\tau \omega }\quad \quad \;d=2\vspace*{0.1in}$ \\ 
%$-(\omega \tau )^{1/2}\quad d=3\vspace*{0.1in}\quad .$%
%\end{tabular}
%\right.  
\label{eq:4.10}
\end{equation}
Note that the low frequency correction to
$\sigma_0$ in Eq.\ ({\ref{eq:4.10}) diverges as $\omega\rightarrow 0$ 
for $d\leq 2$. Further, these contributions have a sign such as to decrease
the conductivity. In a more complete theory that effectively resums all such
divergent contributions, $\sigma$ is identically equal to zero for 
$\omega = 0$ in $d\leq 2$ \cite{LeeRama}. This phenomenon is known as
Anderson localization. In $d>2$ a metal-insulator transition,
called the Anderson transition, occurs as a function of the disorder
or the Fermi energy. In $d=2$ there cannot be a metallic phase, as the
electrons are always localized. $d=2$ is therefore the lower 
critical dimension for the Anderson transition.

The same results can also be obtained from the generalized Boltzmann
equation derived in Sec. 3. Resumming all of the backscattering or MCD
diagrams shown in Fig.\ \ref{fig:3} leads to Eq.\ (\ref{eq:4.10}) 
with explicit coefficients. The
crucial point is that the summations of the MCD can be related, in the
absence of a magnetic field, to the summation of ladder diagrams that give
$\chi_{nn}$. Analytically, the MCD lead to a contribution of the form
\begin{equation}
\frac{\delta \sigma }{\sigma _{0}} \approx -c_{d}\int\limits_{q<\ell^{-1}}
       \frac{1}{-i\omega +D_{\rm B}q^{2}}\quad.
\label{eq:4.11}
\end{equation}
Here ${\bf q}={\bf p}+{\bf p}_{1}$ is the backscattering
momentum, a long-wavelength approximation has been made so that $q<\ell
^{-1},$ and $c_{d}\sim \lambda_{\rm F}^{d-1}$ is a constant. Performing
the integral gives Eq.\ (\ref{eq:4.10}). 
At finite temperatures the $\omega $ in Eq.\ (\ref{eq:4.10}) is effectively
replaced by $T$ (see below), 
so that $\sigma$ depends nonanalytically, and in $d=1,2$
singularly, on $T$. In Fig.\ \ref{fig:7}, the resistivity
$\rho \sim 1/\sigma$ is plotted versus $\ln 1/T$
for a two-dimensional amorphous film. The straight line confirms the
existence of the logarithm in Eq.\ (\ref{eq:4.10}).
\begin{figure}[ht]
\centerline{\psfig{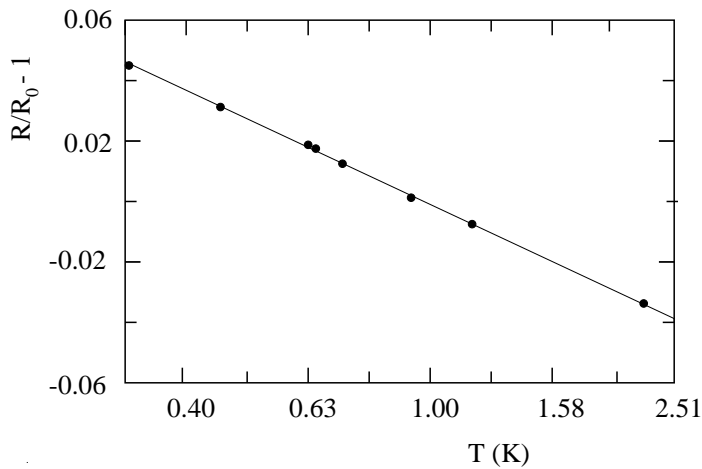}\vspace*{5mm}}
\caption{Resistance, $R$, normalized to $R_0 = R(T=1K)$,
 of a thin PdAu film plotted versus $\log T$ as measured by Dolan and
 Osheroff \protect\cite{DolanOsheroff}. From Ref.\ \protect\cite{us_ernst}.}
\label{fig:7}
\end{figure}

It is worth noting that the hydrodynamic pole in Eq.\ (\ref{eq:4.11}) 
occurs because the MCD are related to a spontaneously broken symmetry and a
corresponding Goldstone mode \cite{SchaferWegner}.
This mode is soft or massless only at zero temperature. That is,
the relationship of the MCD to $\chi_{\rm nn}$ that was noted above, 
is valid only for $T=0$.
At finite temperature, the $\omega$ in Eq.\ (\ref{eq:4.11}) 
is replaced by $\omega + 1/\tau_{\rm in}$, with
$\tau_{\rm in}$ the temperature dependent inelastic mean-free time. 
The Goldstone modes thus
acquire a mass, in contrast to $\chi_{\rm nn}$ 
which is always massless due to particle number
conservation. In the literature, in analogy to particle physics terminology,
these modes that are hydrodynamic only at
$T=0$ are often called either `diffusons' or `Cooperons', 
depending on how they arise.

\subsection{The effects of electron-electron interactions}
\label{subsec:IV.C}

In the absence of electron-electron interactions it is easy to show that
thermodynamic quantities do not depend on the dynamical diffusive processes
that lead to the low frequency nonanalyticities discussed in the previous
subsection. However, when interactions are included this is no longer the
case, and the static and dynamical properties of any quantum system are
coupled together. From a technical point of view this is obvious 
within a many-body Matsubara formalism. From a scaling viewpoint, in 
fundamental units $%
(\hbar =1=k_{\rm B})$ $T$ and $\omega $ have the same units, implying
thermodynamics $(T)$ and dynamics $(\omega )$ are coupled together.

To motivate the static results, we note that for diffusive systems, $\omega $
scales like $k^{2}.$ For the wavenumber and frequency dependent
conductivity this implies, for $2<d<4,$
%\begin{mathletters}
\label{eqs:4.12}
\begin{equation}
\frac{\sigma (k=0,\omega)}{\sigma_{0}} = 1 + a\,\omega^{(d-2)/2}+\ldots\quad, 
\label{eq:4.12a}
\end{equation}
\begin{equation}
\frac{\sigma (k,\omega=0)}{\sigma_{0}} = 1 + a^{\prime}\,k^{d-2}+\ldots \quad .
\label{eq:4.12b}
\end{equation}
%\end{mathletters}%
Note that the coefficients $a$ and $a^{\prime }$ in Eqs.\ (\ref{eqs:4.12})
are positive since wave interference effects decrease $\sigma (0,0)$. Another
interesting quantity is the static spin susceptibility, $\chi_{s}(k)$, 
as a function of the wavenumber. Disorder slows down the electrons, so that
in effect any two electrons spend more time close to one another, i.e., disorder
effectively increases the strength of the electron-electron interactions.
This in turn implies a disorder induced increase in $\chi _{s}$.
Calculations similar to those that lead to Eqs.\ (\ref{eq:4.12a}),
(\ref{eq:4.12b}) give, at $T=0$ \cite{us_chi_s},
\begin{equation}
\frac{\chi _{s}(k)}{\chi _{s}(0)} = 1-\widetilde{a}\,k^{d-2}+0(k^{2})\quad .
\label{eq:4.13}
\end{equation}
In real space, this in turn implies long-range equal-time spin correlations
that decay like
\begin{equation}
\chi _{s}(r)\sim 1/r^{2(d-1)}\quad .  
\label{eq:4.14}
\end{equation}
This is a remarkable result. In classical statistical mechanics, $%
\chi _{s}(r)\propto \exp (-r/\xi)$, with $\xi $ the correlation length which,
away from any critical point,
is on the order of the lattice spacing. In
contrast, at $T=0$, the correlations in Eq.\ (\ref{eq:4.13}) are power-law like 
everywhere in the phase diagram.
These long-range correlations are due to a
coupling of the hydrodynamic spin transport to the charge diffusion 
processes which themselves are intrinsically of long range.

Physically, Eq.\ (\ref{eq:4.13}) 
implies that spin fluctuations effectively interact
between themselves through long-range interactions. It is well known that
long-range interactions have a profound effect on phase transitions and
critical behavior. The implication of Eq.\ (\ref{eq:4.13}) for the
paramagnetic to ferromagnetic quantum phase transition is discussed in the
following lectures \cite{DB's_lecture}.
Note that long-range spatial correlations also exist in
{\it nonequilibrium classical} systems. 
In that case, statics and dynamics are also
coupled together, somewhat analogous to the {\it equilibrium quantum} 
statistical mechanical systems discussed here \cite{us_ernst}.

\section{Final Remarks}

We conclude with a few additional remarks.
Although we have not emphasized this fact, the interesting phenomena in quantum
kinetic theory discussed above have strong analogies in classical
kinetic theory. Indeed, the discovery of the logarithmic
singularity in the disorder expansion of $\sigma $, Eq.\ (\ref{eq:4.3}), 
was motivated by much earlier, related
considerations in classical kinetic theory \cite{DKS}. 
Similarly, long-time tail
effects and low-frequency nonanalyticities were discussed already in 1967 
for classical systems \cite{AlderWainwright}, 
whereas the weak localization effects discussed above
were not discovered until 1979 \cite{AALR}.

Another aspect we have not stressed is that, even though the phenomena in
classical and quantum systems are similar, the correlations in space and
time are stronger in the quantum mechanical case. For example, in the
classical Lorentz gas the low frequency nonanalytic contribution to 
the conductivity is proportional to $\omega^{d/2}$. Physically,
this eventually leads to the
conclusion that while classical particles are diffusive in $d=2$, there is
no diffusive phase in two-dimensional quantum systems \cite{AALR,LeeRama}.

Finally, one other important feature in quantum systems is the coupling
between statics and dynamics. In classical systems the thermodynamic
correlation functions do not couple to the dynamical fluctuations that lead
to the long-range correlations in time correlation functions. In zero
temperature quantum systems the situation is different. As explained in the
last section, equal time correlations do couple to these dynamical
fluctuations, and as a consequence of this, there are both long-ranged, 
power-law spatial correlations and long-ranged, power-law time correlations in
quantum systems.

\acknowledgements

This work was supported by the NSF under grant numbers DMR-96-32978 and
DMR-98-70597. This research was supported in part by the National Science
Foundation under grant No. PHY94-07194.


\begin{thebibliography}{2}
\bibitem{AshcroftMermin} See, e.g.,  N.W. Ashcroft and N.D. Mermin, 
 {\it Solid State Physics} (Holt, Rinehart and Winston, New York 1976).
\bibitem{UU} E.A. Uehling and G. Uhlenbeck, Phys. Rev. {\bf 43}, 552 (1933).
\bibitem{Edwards} S.F. Edwards, Philos. Mag. {\bf 3}, 1020 (1958).
\bibitem{AGD} A.A. Abrikosov, L.P. Gorkov, and I.E. Dzyaloshinski, {\it Methods
 of Quantum Field Theory in Statistical Physics} (Dover, New York 1975).
\bibitem{FetterWalecka} See, e.g., A.L. Fetter and J.D. Walecka, {\it Quantum
 Theory of Many-Particle Systems} (McGraw-Hill, New York 1971).
\bibitem{Kubo} R. Kubo, J. Phys. Soc. Japan {\bf 12}, 570 (1957); see also
 G.D. Mahan, {\it Many-Particle Physics} (Plenum, New York 1981), ch. 3.7.
\bibitem{VW} D. Vollhardt and P. W{\"o}lfle, Phys. Rev. B {\bf 22}, 4666 (1980).
\bibitem{density_expansion} T.R. Kirkpatrick and J.R. Dorfman, J. Stat. Phys. 
 {\bf 30}, 67 (1983); T.R. Kirkpatrick and J.R. Dorfman, Phys. Rev. A {\bf 28},
 1022 (1983); T.R. Kirkpatrick and D. Belitz, Phys. Rev. B {\bf 34}, 2168
 (1986).  
\bibitem{DKS} For a review see, e.g., J.R. Dorfman,
 T.R. Kirkpatrick, and J.V. Sengers, Ann. Rev. Phys. Chem. {\bf 45}, 213
 (1994).
\bibitem{2d_density_expansion} F. Evers, D. Belitz, and W. Park, Phys. Rev.
 Lett. {\bf 78}, 2768 (1997). 
\bibitem{us_PRE} K.I. Wysokinski, W. Park, D. Belitz, and T.R. Kirkpatrick, 
 Phys. Rev. E {\bf 52}, 612 (1995).
\b{Adams} P.~W. Adams, D.~A. Browne, and M.~A. Paalanen, Phys. Rev. B
 {\bf 45}, 8837 (1992).
\bibitem{us_ernst} For a review, see, e.g., T.R. Kirkpatrick and D. Belitz,
 J. Stat. Phys. {\bf 87}, 1307 (1997).
\bibitem{LeeRama} For a review, see, e.g., P.A. Lee and T.V. Ramakrishnan, 
 Rev. Mod. Phys. {\bf 57}, 287 (1985).
\bibitem{Bergmann} G. Bergmann, Phys. Rep. {\bf 101}, 1 (1984).
\b{DolanOsheroff} G.~J. Dolan and D.~D. Osheroff, Phys. Rev. Lett.
 {\bf 43}, 721 (1979).
\bibitem{SchaferWegner} L. Sch{\"a}fer and F. Wegner, Z. Phys. B {\bf 38}, 113 
 (1980).
\bibitem{us_chi_s} D. Belitz, T.R. Kirkpatrick, and T. Vojta, 
 Phys. Rev. B {\bf 55}, 9452 (1997).
\bibitem{DB's_lecture} D. Belitz and T.R. Kirkpatrick, following paper,
 cond-mat/98xxxxx.
\bibitem{AALR} E. Abrahams, P.W. Anderson, D.C. Licciardello, and 
 T.V. Ramakrishnan, Phys. Rev. Lett. {\bf 42}, 673 (1979).
\b{AlderWainwright} B.J. Alder and T.E. Wainwright, Phys. Rev. Lett. {\bf 18},
 988 (1967).
\end{thebibliography}
\end{document}